\documentclass[iop]{emulateapj}
\usepackage{amsmath}
\usepackage{natbib,url}
\usepackage{nicefrac}
\usepackage{subfigure}
\usepackage{hyperref}
\hypersetup{colorlinks=true, urlcolor=blue, citecolor=blue}

\usepackage{color}
\usepackage{soul}

\usepackage{textcomp}
\usepackage{amssymb}
\shorttitle{Multiwavelength Observations of XTE J1752-223}
\shortauthors{Chun et al.}

\submitted{accepted for publication in ApJ}

\usepackage{graphicx}

\newcommand{\rxte}{\emph{RXTE}}

\newcommand{\xmm}{\emph{XMM}}

\newcommand{\integral}{\emph{INTEGRAL}}

\newcommand{\wsim}{\ensuremath{\sim}}

\newcommand{\ST}{XTE~J1752$-$223}
\newcommand{\swift}{\emph{Swift}}

\begin{document}

\title{Multiwavelength observations of the black hole transient XTE J1752-223 during its 2010 outburst decay}

\author{Y. Y. Chun\altaffilmark{1},
        T. Din{\c c}er\altaffilmark{1},
        E. Kalemci\altaffilmark{1},
        T. G{\" u}ver\altaffilmark{1},
        J. A. Tomsick\altaffilmark{2},
        M. M. Buxton\altaffilmark{3},
        C. Brocksopp\altaffilmark{4}, \\
        S. Corbel\altaffilmark{5} and
        A. Cabrera-Lavers\altaffilmark{6}
       }

\altaffiltext{1}{Sabanci University, Faculty of Engineering and Natural Sciences, Orhanli, Istanbul, 34956, Turkey}

\altaffiltext{2}{Space Sciences Laboratory, 7 Gauss Way, University of
California, Berkeley, CA, 94720-7450, USA}

\altaffiltext{3}{Yale University, New Haven, CT,  USA}

\altaffiltext{4}{Mullard Space Science Laboratory, University College London, Holmbury St Mary, Dorking, Surrey RH5 6NT, UK}

\altaffiltext{5}{AIM - Unit\'e Mixte de Recherche CEA - CNRS - Universit\'e Paris VII - UMR 7158, CEA Saclay, Service d'Astrophysique, F-91191 Gif sur Yvette, France}

\altaffiltext{6}{Instituto de Astrofisica de Canarias, E-38205 La Laguna, Tenerife, Spain}

\begin{abstract}

Galactic black hole transients show many interesting phenomena during outburst decays. We present simultaneous X-ray (\emph{RXTE}, \emph{Swift}, and \emph{INTEGRAL}), and optical/near-infrared (O/NIR) observations (SMARTS) of the X-ray transient XTE~J1752$-$223 during its outburst decay in 2010. The multiwavelength observations over 150 days in 2010 cover the transition from soft to hard spectral state. We discuss the evolution of radio emission is with respect to the O/NIR light curve which shows several flares. One of those flares is bright and long, starting about 60 days after the transition in X-ray timing properties. During this flare, the radio spectral index becomes harder. Other smaller flares occur along with the X-ray timing transition, and also right after the detection of the radio core. We discuss the significances of these flares. Furthermore, using the simultaneous broadband X-ray spectra including \emph{INTEGRAL}, we find that a high energy cut-off with a folding energy near 250 keV is 
necessary around the time that the compact jet is forming. The broad  band spectrum can be fitted equally well with a Comptonization model. In addition, using photoelectric absorption edges in the \emph{XMM-Newton} RGS X-ray spectra and the extinction of red clump giants in the direction of the source, we find a lower limit on the distance of $>$ 5 kpc. 

\end{abstract}


\section{Introduction}\label{sec:intro}

Galactic black hole transients (hereafter GBHTs) show distinct spectral and temporal properties during the whole outburst across all wavelengths. According to their X-ray spectral and timing properties, these systems are found mainly in the soft state (SS) or in the hard state (HS). Here, we provide a brief summary of properties of GBHTs in these states \citep[see][for details]{McClintock06book,Belloni10}. In the SS, the X-ray emission is dominated by soft photons from a geometrically thin, optically thick disk, which  is modeled by a multi-temperature blackbody \citep{Makishima86}. The variability is low in the SS,  often below the detection limit of \rxte\ for short observations. In the HS, the flux from the thin disk is weak, and the emission is dominated by a hard component, phenomenologically modeled with a power-law in the X-ray spectrum. The variability is strong (typically $>$20\% rms amplitude),  and the power spectra seen in this state can exhibit a variety of features such as quasi-periodic 
oscillations. There also exist intermediate states (IS), usually observed when the source makes a transition between SS and HS. 

The emission properties in radio, optical and infrared bands also strongly depend on the X-ray spectal states. The radio emission
is quenched in the SS \citep{Fender99,Corbel02, Russell11}.  In the HS, radio observations indicate compact steady jets \citep{Fender01b}, and synchrotron radiation from the compact jet has also been detected in optical, and near infrared \citep{Buxton04, Kalemci04, Migliari07, Coriat09, Buxton12, Kalemci12}. For XTE~J1550$-$564,  it has been reported that the direct jet synchrotron emission may dominate the entire X-ray emission in the hard state at $10^{-4}\,<L_{Edd}<10^{-3}$ \citep{Russell10}. Similar analysis indicated that the jet producing significant X-ray flux at low luminosities is also a possibility for XTE~J1752$-$223 \citep{Russell12}.

During outburst decays, these sources display state transitions from the SS to IS to HS. Since the jet emission is quenched in the SS, multiwavelength observations of outburst decays allow us to investigate the X-ray spectral and timing properties required for jet formation and to constrain the jet's contribution to the X-ray emission. 

A possible effect of the jet on the X-ray spectral properties is the injection of non-thermal electrons in the corona, thereby hardening the X-ray spectrum. A cut-off in high energy spectrum of these sources is often observed in the hard state, usually interpreted as a sign of thermal Comptonization. Sometimes the cut-off disappears in high energies after jets are observed as in 4U 1543$-$47 \citep{Kalemci05}, and GRO J1655$-$40 \citep{Kalemci06_mqw, Caballero07}). However, there are sources that do not require a cut-off for the entire outburst (H1743$-$322 \citealt{Kalemci06}, GX 339-4 during its decay in 2005 \citealt{Kalemci06_mqw}, XTE J1720$-$318 \citealt{CadolleBel04}). \cite{Miyakawa08} investigated the presence of cut-off from all bright hard state observations of GX 339-4 observed with HEXTE on \rxte, yet the statistics were not good enough to constrain the evolution of the cut-off parameters. Utilizing \integral\ ISGRI data in addition to the HEXTE data provided information about the evolution of 
the cut-off from the hard state to the  hard intermediate state in the rising phase of the outburst of GX 339-4 \citep{Motta09}. However, even in the brighter outburst rise, the relation between high energy cut-off parameters and presence of jets is not well established. For example, with the same data set but with a slightly different set of instruments, \cite{Caballero07} and \cite{Joinet08} reached conflicting results about the presence of cut-offs in the hard state spectrum of GRO J1655-40 during the outburst rise while the jet emission was present.

Measuring distances of Galactic black hole sources provides important information about the birthplaces of X-ray binaries, and, more importantly, allows us to constrain the luminosities of these sources. Obtaining the X-ray luminosity of these sources would allow us to tie spectral states to physical processes in the accretion physics. It was shown by \cite{Maccarone03_b} that transition luminosities occur at similar values. This can be related to the changes from an accretion regime dominated by the disk to one dominated by a corona. Furthermore, this could also help in determining if there is a critical luminosity at which jets can form, and if jet dominated states exist at low luminosities \citep[e. g.][]{Russell10}, at what luminosity this takes place.  Moreover, luminosities are important for studying the fundamental plane that connects the radio - X-ray correlation \cite{Corbel00,Corbel03} from stellar mass black holes to the supermassive black holes \citep{Merloni03,Falcke04,Kording06} as well as the 
variability studies that connect these black hole sources \citep{Kording07}. 

\subsection{XTE J1752-223}\label{subsec:1752}

XTE J1752-223 was discovered in the Galactic bulge region with \emph{RXTE} on 23 October 2009 \citep{Markwardt09a}. It was suggested to be a black hole candidate by \cite{Markwardt09b}. Strong, relativistic iron emission lines are detected by \emph{Suzaku} and \emph{XMM-Newton} \citep{Reis11}. The source showed typical outburst evolution of GBHTs, and it was intensely monitored using several satellites: \emph{RXTE} \citep{Munoz10, Shaposhnikov10_atel}, \emph{MAXI} \citep{Nakahira10} and \emph{Swift} \citep{Curran11}. A distance of 3.5$\pm$0.4 kpc and black hole mass of 8--11\emph{M$_{\odot}$} were determined by \cite{Shaposhnikov10} using QPO frequency saturation and comparison with other sources. \cite{Ratti12} recently discussed the X-ray -- radio correlation in this source. They also suggested that the source is in the Galactic bulge or closer to us, i.e. $<$ 8 kpc based on comparison of transition luminosities of black hole transients \citep{Maccarone03_b}. The source's core location was accurately 
determined using the astrometric optical observations and the VLBI radio imaging \citep{MillerJ11} in addition to the detection of decelerating jets and receding ejecta \citep{Yang11}.

\begin{figure}
\epsscale{1.2}
\plotone {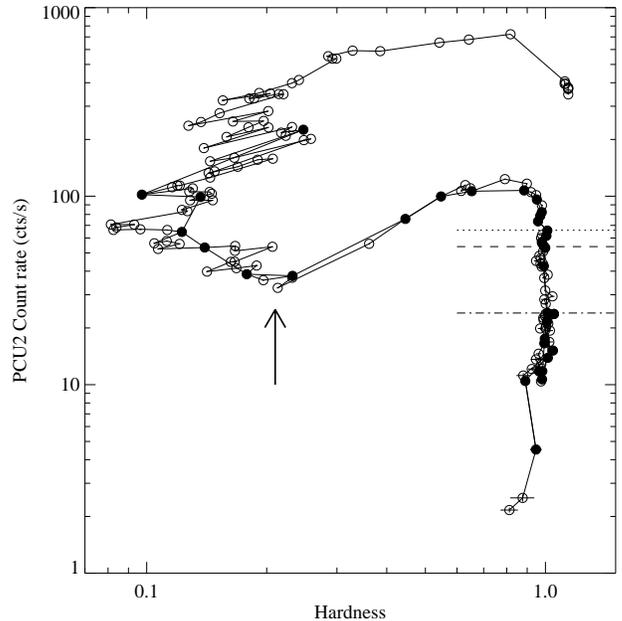}
\caption {Hardness-intensity diagram of XTE J1752-223. Only \rxte\ PCA PCU2 data were used. Y-axis is count rate in 2.8 - 25 keV. Hardness is defined as the ratio of  (5.2 - 9.8 keV) to (2.8 - 5.2 keV) band count rates. The filled circles are the observations used in this work. The vertical arrow show the hardness/intensity at the timing transition. The dotted line show the hardness/intensity during the \emph{INTEGRAL} observation.  The dashd line show the hardness/intensity at the VLBI detectıon of the core.  The dot-dash line show the hardness/intensity at the near infrared peak when the ATCA radio spectrum is consistent with optically thick emission.
}
\label{fig:hid}
\end{figure}
\epsscale{1.0}

\cite{Russell12} discussed a late jet re-brightening in the HS via multiwavelength observations during its outburst decay towards quiescence (this is called flare 3 in our paper, see Appendix). They suggested that the brightening in the optical and X-rays probably has a common origin, but it is not clear if the origin is synchrotron radiation from a compact jet. They also suggested that the jet power changes at the peak of the flare and that there is evidence that the jet break between optically thin and thick synchrotron emission shifts to higher frequencies during this flare.

In this paper, we present an in-depth multiwavelength analysis of XTE J1752-223 during its outburst decay, covered well with \rxte\, and \emph{Swift} in X-rays, SMARTS in O/NIR, and \emph{Australia Telescope Compact Array} (ATCA) in radio. The spectral and temporal analyses for the source were examined for the whole outburst decay by simultaneous monitoring with \emph{RXTE} and \emph{Swift} during MJD 55,240--55,390. Fig.~\ref{fig:hid} shows the hardness-intensity diagram of the source during the entire outburst obtained using the \rxte\ data. We report on the observations for which simultaneous {\em Swift} data are available which are shown with filled circles in Fig.~\ref{fig:hid}. The X-ray behavior of the source is compared to the evolution in NIR and optical as well as the radio. The decaying part of the outburst was covered previously with multiwavelength studies by \cite{Russell12} and \cite{Brocksopp13}. However, our work enhances previous studies in several ways. We show for the first time the SMARTS light curve in the optical and the near infrared, which allows us to investigate the relation between the radio emission, X-ray emission and optical/NIR emission in much more detail than before. In addition, we use \emph{RXTE} and \emph{Swift} observations together for X-ray fits, and model the effects of Galactic ridge emission to obtain reliable spectral and temporal parameters at low flux levels. The X-ray timing properties of the source and their relation to the multiwavelegth emission are investigated in terms of both evolution of rms amplitudes and the shape of the power spectra. Moreover, an additional 3-day (MJD 55,305--55,307) broadband X-ray spectrum (\emph{Swift}, \rxte, and \integral), covering 0.6 -- 200 keV, was also obtained in order to constrain the high energy behavior of the source (the time of the \integral\ obsrvation is shown in Figs.~\ref{fig:hid}, ~\ref{timing1752}, and ~\ref{opticalra}). Finally, we analyze the \emph{XMM-Newton} RGS spectra to determine the Hydrogen column density ($N_{H}$) of the source and use this information to determine the distance. We compare the result with the previously reported distance estimates \citep{Shaposhnikov10, Ratti12}.


\begin{figure}
\epsscale{1.2}
\plotone {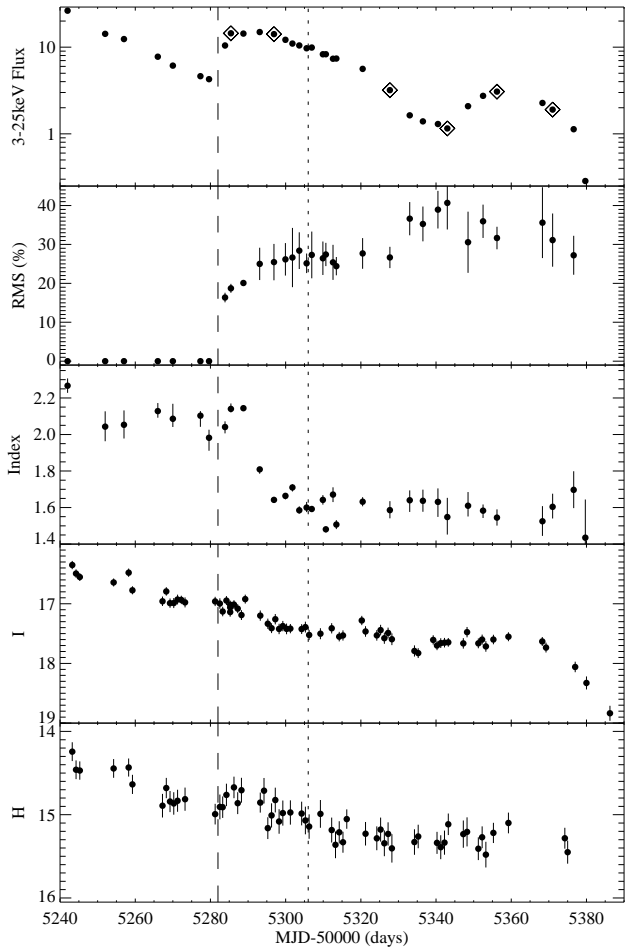}
	\caption {The evolution of spectral and temporal properties of XTE J1752-223. The photon index was obtained from combined X-ray spectra of XRT and PCA. The rms and flux after MJD~55,303 were corrected due to the ridge emission in PCA. The fast transition determined by timing analysis is indicated by a dashed line at MJD~55,282. The dotted line shows the time of the \emph{INTEGRAL} observation. The two panels on the bottom are of the $I$ and $H$ band NIR light curves, respectively. The observations shown in diamonds in the top panel correspond to those for which we show the power spectra in Fig.\ref{timing1752}
} \label{evolution}

\end{figure}
\epsscale{1.0}

\begin{deluxetable*}{ccccccccc}
\tablewidth{0pt}
\tablecaption{Log of X-ray spectral and temporal parameters}
\tablehead{
\colhead{MJD} & \colhead{XTE ObsId} & \colhead{Swift ObsId\tablenotemark{a}} & \colhead{Flux\tablenotemark{b}}  & \colhead{rms (\%)} & \colhead{$\Gamma$} & \colhead{$T_{in} (keV)$} & \colhead{$\nu_{1}$ ($Hz$)\tablenotemark{c}} & \colhead{note} }
\startdata
55242&95360-01-04-04&31532014&26.328&\dots&2.27$_{-0.04}^{+0.04}$&0.57$_{-0.01}^{+0.01}$&\dots&\tabularnewline
55252&95360-01-05-02&31532021&14.236&\dots&2.04$_{-0.08}^{+0.08}$&0.53$_{-0.01}^{+0.01}$&\dots&\tabularnewline
55257&95360-01-06-04&31640001&12.397&\dots&2.05$_{-0.08}^{+0.08}$&0.52$_{-0.01}^{+0.01}$&\dots&\tabularnewline
55266&95360-01-08-00&31640003&7.7581&\dots&2.13$_{-0.04}^{+0.04}$&0.49$_{-0.01}^{+0.01}$&\dots&\tabularnewline
55270&95360-01-08-03&31640005&6.1194&\dots&2.09$_{-0.05}^{+0.08}$&0.48$_{-0.01}^{+0.01}$&\dots&\tabularnewline

55277&95360-01-09-03&31640006&4.6238&\dots&2.10$_{-0.06}^{+0.03}$&0.48$_{-0.01}^{+0.01}$&\dots&\tabularnewline
55580&95360-01-09-05&31640007&4.2723&\dots&1.98$_{-0.07}^{+0.04}$&0.49$_{-0.01}^{+0.01}$&\dots&\tabularnewline
55284&95360-01-10-02&31640008&10.445&16.36$\pm1.24$&2.04$_{-0.04}^{+0.03}$&0.45$_{-0.01}^{+0.01}$&3.15$\pm0.68$&\tabularnewline
55285&95360-01-10-04&31640009&14.444&18.72$\pm1.15$&2.14$_{-0.02}^{+0.03}$&0.41$_{-0.01}^{+0.01}$&\bf{1.83}$\pm0.58$&\tabularnewline
55289&95360-01-11-00&31640010&14.334&20.09$\pm0.68$&2.14$_{-0.02}^{+0.01}$&0.42$_{-0.01}^{+0.01}$&1.67$\pm0.24$&\tabularnewline

55293&95360-01-11-04&31640011&14.924&25.00$\pm4.15$&1.81$_{-0.02}^{+0.01}$&0.29$_{-0.01}^{+0.02}$&0.55$\pm0.22$&\tabularnewline
55297&95360-01-12-01&31640012&14.129&25.46$\pm4.68$&1.64$_{-0.01}^{+0.01}$&0.31$_{-0.04}^{+0.04}$&\bf{0.35}$\pm0.12$&\tabularnewline
55300&95360-01-12-03&31688001&12.157&26.17$\pm4.17$&1.66$_{-0.02}^{+0.01}$&0.25$_{-0.02}^{+0.03}$&0.36$\pm0.13$&\tabularnewline
55302&95360-01-12-04&31688003&10.993&26.63$\pm7.61$&1.71$_{-0.02}^{+0.02}$&0.26$_{-0.02}^{+0.03}$&0.41$\pm0.10$&\tabularnewline
55304&95702-01-01-01&31688006&10.462&28.40$\pm4.72$&1.59$_{-0.02}^{+0.02}$&0.29$_{-0.03}^{+0.03}$&0.35$\pm0.10$&ISGRI\tabularnewline

55306&95702-01-01-03&31688008&9.7074&25.15$\pm2.51$&1.60$_{-0.03}^{+0.03}$&0.25$_{-0.05}^{+0.10}$&0.41$\pm0.17$&ISGRI\tabularnewline
55307&95702-01-01-04&31688009&9.8888&27.30$\pm6.00$&1.59$_{-0.02}^{+0.02}$&0.25$_{-0.02}^{+0.02}$&0.29$\pm0.11$&\tabularnewline
55310&95702-01-02-00&31640013&8.2724&26.44$\pm4.30$&1.64$_{-0.03}^{+0.03}$&0.31$_{-0.05}^{+0.08}$&0.33$\pm0.13$&\tabularnewline
55311&95702-01-02-01&31640014&8.2815&27.41$\pm2.99$&1.48$_{-0.02}^{+0.01}$&0.27$_{-0.03}^{+0.04}$&0.41$\pm0.13$&\tabularnewline
55313&95702-01-02-03&31688012&7.3498&25.40$\pm4.51$&1.67$_{-0.04}^{+0.04}$&0.25$_{-0.04}^{+0.07}$&0.43$\pm0.26$&\tabularnewline

55313.5&95702-01-02-04&31688013&7.3846&24.41$\pm2.35$&1.51$_{-0.03}^{+0.02}$&0.33$_{-0.06}^{+0.16}$&0.32$\pm0.09$&\tabularnewline
55320.5&95702-01-03-04&31688014&5.6153&27.69$\pm3.95$&1.63$_{-0.03}^{+0.03}$&No disk&0.36$\pm0.12$&\tabularnewline
55328&95702-01-04-04&31688016&3.1901&26.65$\pm2.74$&1.59$_{-0.05}^{+0.05}$&needed&\bf{0.29}$\pm0.11$&a single\tabularnewline
55333&95702-01-05-03&31688017&1.6355&36.62$\pm4.26$&1.64$_{-0.06}^{+0.05}$&&0.17$\pm0.06$&Lorentzian\tabularnewline
55336.5&95702-01-05-06&31688018&1.3928&35.26$\pm4.47$&1.64$_{-0.06}^{+0.06}$&&0.15$\pm0.05$&enough\tabularnewline

55340.5&95702-01-06-02&31688019&1.2990&38.94$\pm4.83$&1.63$_{-0.08}^{+0.07}$&&0.15$\pm0.04$&\tabularnewline
55343&95702-01-06-03&31688020&1.1554&40.67$\pm6.84$&1.55$_{-0.10}^{+0.11}$&&\bf{0.22}$\pm0.10$\tabularnewline
55348.5&95702-01-07-05&31688021&2.0856&30.56$\pm7.86$&1.61$_{-0.06}^{+0.08}$&&$\leq$1.41&\tabularnewline
55352.5&95702-01-08-00&31688022&2.7439&35.94$\pm4.28$&1.58$_{-0.04}^{+0.03}$&&0.25$\pm0.05$&\tabularnewline
55356&95702-01-08-02&31688023&3.0706&31.65$\pm2.90$&1.55$_{-0.04}^{+0.05}$&&\bf{0.19}$\pm0.07$\tabularnewline

55368&95702-01-10-01&31688024&2.2719&35.59$\pm9.11$&1.53$_{-0.08}^{+0.08}$&&0.20$\pm0.12$&\tabularnewline
55371&95702-01-10-02&31688025&1.9037&31.11$\pm6.84$&1.60$_{-0.06}^{+0.07}$&&\bf{0.18}$\pm0.10$\tabularnewline
55377&95702-01-11-01&31688026&1.1313&27.21$\pm5.00$&1.70$_{-0.10}^{+0.10}$&&0.09$\pm0.04$&\tabularnewline
55380&95702-01-12-00&31688027&0.2859&...&1.44$_{-0.21}^{+0.21}$&&4.18$\pm1.74$&\tabularnewline
\enddata
\tablenotetext{a}{Spectral parameters are obtained from the joint \rxte\ PCA and \swift\ XRT data fits}
\tablenotetext{b}{The unabsorbed flux values between 3--25 keV are in units of 10$^{-10}$ ergs/cm$^{2}$/s.}
\tablenotetext{c}{The timing properties of observations. Those shown in bold are discussed further in text and in Fig.~\ref{timing1752}. $\nu_1$ stands for the peak frequency of the Lorentzian that peaks at lower frequencies. }
 
 \label{journal}
\end{deluxetable*}

\section{Data Reduction}\label{sec:data}

\subsection{\rxte}\label{subsec:rxte}

The X-ray evolution of XTE J1752--223 in 2010 outburst is shown in the top three panels of Fig.~\ref{evolution}: unabsorbed flux in 3-25 keV, rms amplitude of variability, and power-law index, respectively. The source was monitored throughout the outburst decay, including the fast transition from SS to HS at MJD~55,282 until it went below detection after MJD~55,380. 

The \rxte\ PCA data were reduced with the scripts developed at UC San Diego and University of T{\"u}bingen using HEASOFT v6.7. In the extraction of the energy spectra, the photons from all available PCUs were considered. The background spectra were created from ``bright'' or ``faint'' models on the basis of the net count rate being greater or less than 70 c/s/PCU, respectively. We added 0.5\% systematic error to all PCA spectra as suggested by the \rxte\ team.

We used T{\"u}bingen Timing Tools in IDL to compute the power density spectra
(PDS) of all observations using PCA light curves in 3-25 keV band. The dead time
effects were removed according to \cite{Zhang95} with a dead-time of $\rm 10\,\mu s$ 
per event, and the PDS is normalized according to \cite{Miyamoto89}. Broad and narrow 
Lorentzians (quality factor greater than 2 are denoted as QPOs) are used for fitting \citep{Kalemci05,Pottschmidt02th}. The rms amplitudes are calculated by integrating rms-normalized PDS from 0 Hz to infinity. The peak frequencies are calculated as described in \cite{Belloni02}.

\subsection{INTEGRAL}\label{subsec:inte}

We observed the source using \integral\ (the INTernational Gamma-Ray Astrophysics Laboratory), observation ID ``07400260001'' for dates between MJD 55,304 and 55,306 (Rev. 0917, 0918). The IBIS and SPI data were reduced and analyzed with the help of the ISDC Off-line Scientific Analysis (OSA) software package 9.0, and only the ISGRI spectra were used for further analysis in order to get better statistics at the relevant energy range. The \integral\ data were divided into three parts with approximately equal observing time to check whether the spectrum evolves at high energies within 2 days. Since the fit parameters did not vary within the errors, we decided to merge two revolutions in order to increase statistics, i.e., the total integration time became 177.4~ks. The observation time is indicated by a dotted line in Fig.~\ref{evolution}. Note that neither spectral nor temporal properties show significant evolution around the time of the \integral\ observation.

\subsection{Swift}\label{subsec:swift}

The \emph{Swift} XRT data were processed with the standard procedures (xrtpipeline v0.12.3). In the
observations we analyzed, Windowed Timing mode data were available before MJD~55,367 and Photon Counting mode data were used afterwards. The energy spectra of the source were extracted from the photons within a circle of radius 35{"} centered at the source position. The background spectra were extracted from photons within regions between 70{"} and 100{"} from XTE J1752-223 centroid. The selection of event grades was 0-2 for Windowed Timing mode, and 0-12 for the Photon Counting mode. XRT response matrices swxwt0to2s6\_20070901v012.rmf and swxpc0to12s6\_20070901v011.rmf were used for the Windowed Timing and Photon Counting modes, respectively. We also generated auxiliary response files with the HEASOFT tool xrtmkarf. In making the ARF files, we used an exposure map produced with the tool xrtexpomap. There was also some pile-up before MJD~55,272 when the count rates were over 100 c/s, and for these observations we excluded the peak region of radius 5{"}.

\subsection{XMM--Newton}\label{subsec:xmm}

The \emph{XMM-Newton} observation of the source was performed on 6 April 2010, MJD~55,292. The total exposure time of the observation was 41.8 ks. Details of this observation and results from the EPIC-pn detector has been discussed in \cite{Reis11}. We here concentrate on the RGS  (Reflection Grating Spectrometer, \citealt{denHerder01}) data, which provide high resolution soft X-ray spectra in the range of 5 to 35 \AA. Our aim is to model the absorption edges in the soft X-rays to determine the hydrogen column density along the line of sight towards \ST. 

We extracted RGS spectra using the {\it rgsproc} within SAS v11.0.0 with the latest calibration files available as of Dec. 2011. We grouped the RGS spectra with {\it  specgroup} tool so that each spectral channel will have at least 100 counts and the instrumental resolution will not be oversampled more than a factor of three.

Given the X-ray flux of the source at the time of the observation, we also checked the RGS observation against any affect of photon pile-up. For this purpose we used the fluxed RGS X-ray spectra, as given by the Browsing Interface for RGS data (BIRD), and checked the flux on each CCD against the limits given in the XMM-Newton User's Handbook\footnote{\url{http://xmm.esac.esa.int/external/xmm_user_support/documentation/uhb/index.html}}. This comparison showed us that even though there was a photon pile-up it was smaller than 2\%. This is also thanks to the fact that the observation was performed with double-node readout for RGS1 and RGS2. Such a small amount of saturation would have only minimal, if any, effect on the depths of individual absorption edges.

\subsection{Optical \& NIR observations}\label{subsec:nir}

\begin {figure}[h]
\epsscale{1.2}
\plotone{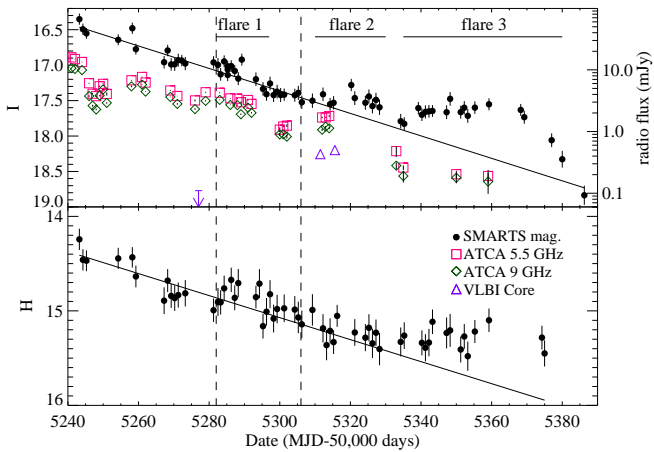}
	\caption {The evolution of SMARTS O/NIR magnitudes and radio fluxes during the decay. The solid lines indicate exponentially decaying emission possibly from the outer disk, which are shown to demonstrate the presence of the three flares mentioned in the text (see Appendix). The radio data are added with diamonds and triangles (see legend). The errors on radio before MJD~55,320 are smaller than the symbol size. The arrow represents the upper limit of the VLBI core radio emission at MJD~55,277 (see text). The dates of the fast transition and the ISGRI observation are indicated by dashed lines on MJD~55,282 and 55,306, respectively. A colored version of this figure is available in the online version.} \label{opticalra}
\end{figure}

In addition to the X-ray observations, we analyzed the O/NIR SMARTS data ($I$ and $H$ band) obtained with the CTIO 1.3m telescope, during MJD 55,240--55,390 covering 150 days of the outburst decay. With the help of IRAF V2.14, we conducted photometry using the PSF (Point Spread Function) fitting due to the crowded region towards the source (see \citealt{Russell12} for the high resolution NIR image). Then, we selected five comparison stars in the FOV as non-variables and went through the standard routines to get the actual magnitudes (see \citealt{Buxton12} for standard procedures).  The magnitudes we obtained could be affected by a nearby source, especially in the $H$ band due to the poor observational conditions and the resolution of the telescope. The results are shown in  Figs.~\ref{evolution} and \ref{opticalra}.

\epsscale{1.0}

\subsection{Radio}\label{subsec:radio}

The ATCA provided the evolution of the source throughout the decay in radio in two bands, 5.5 and 9 GHz. The peak fluxes in two bands were obtained with the help of the radio interferometry data reduction package, \emph{Miriad} \citep{Sault95}. Details of the analysis and an extended discussion of the behavior of ths source in radio for the entire outburst can be found in \cite{Brocksopp13}. In addition, very long baseline interferometry (VLBI) measurements were also conducted for a few observations during the decay with the European VLBI Network (EVN) and the Very Long Baseline Array.  The high resolution of VLBI made it possible to resolve a compact jet core on MJD~55,311 and 55,315 \citep{MillerJ11} (X-ray hardness/intensity at the time of VLBI detections are shown in Fig~\ref{fig:hid}). We used the radio flux values reported by \cite{Yang11} for the core (denoted as component D in their paper). In Fig.\ref{opticalra}, we also show the 3 $\sigma$ upper limit of $0.11 mJy$ for the core on MJD~55,277  based on the EVN image sensitivity (1 sigma = 0.03 mJy/beam in Table 1 of \citealt{Yang11}) and \wsim80\% (typical value) phase-referencing coherence loss (Jun Yang, personal communication).

\section{Analysis and Results}\label{sec:results}

All {\emph RXTE} and {\emph Swift} spectra were fitted with a phenomenological model of disk blackbody (diskbb) plus power law with absorption. A smeared edge model (smedge, \citealt{Ebisawa94}) was also added to the \rxte\ spectrum to improve the $\chi^{2}$ similar to previous work by this group \citep{Kalemci04, Kalemci05, Kalemci06}.  For photoelectric absorption, we used Tuebingen-Boulder ISM absorption (\emph{TBabs}) with the abundance values of \cite{Wilms00} and the cross sections of \cite{Verner96}. This choice is discussed in more detail in Section \ref{sec:distance}. In addition, due to the proximity of the source to the Galactic plane (\emph{b} = +2$^\circ.$11), the Galactic ridge contribution is significant for \rxte\ when the source is at low luminosity levels. To take into account the Galactic ridge emission, an additional power-law was added to the spectral fit after MJD~55,310, fixing the index to 2.1 and normalization to $1.209\times 10$$^{-2}$ based on values given in \cite{Revnivtsev03}. {\
bf{The ridge emission count rates obtained with this model is consistent with the background rates for the PCA Galactic bulge scans\footnote{\url{http://asd.gsfc.nasa.gov/Craig.Markwardt/galscan/html/XTE_J1752-223.html}}.}}The rms amplitudes were corrected for the ridge contribution after MJD~55,303 as well as for background emission following \cite{Berger94}. 

\subsection{Multiwavelength evolution}\label{mulevol}

Fig.~\ref{evolution} shows the evolution of the X-ray spectral and temporal properties as well as the evolution in O/NIR magnitudes. The vertical line on MJD~55,282 indicates the transition in timing due to the fast change in the rms. An increase in the power-law flux accompanies the increase in the rms amplitude of variability. The X-ray flux continued to increase for a couple of days after the transition, whilst the photon index remained the same until the peak in X-ray flux. As the X-ray flux started to decay, the photon index abruptly decreased to $\sim$1.7. The spectrum continued to harden and the index leveled off around 1.6. A secondary peak was also observed in X-rays after MJD~55,340.

The SMARTS optical and NIR observations showed interesting behavior as shown in Figs.~\ref{evolution} and \ref{opticalra}. Some flares are present in the light curves at around MJD~55,280 (flare 1), MJD~55,320 (flare 2) followed by a large increase in flux starting around MJD~55,335 (flare 3) coinciding with the increase in the X-ray flux \citep{Russell12}. The small increase during flare 1 in the $H$ band coincides with the timing transition which is also reported in \cite{Buxton10atel}. We note that after analyzing the data in the following days, it is found that the increase is not significantly above the continuum as shown in Fig.~\ref{opticalra}. Flare 2 starting at MJD~55,320 in the $I$ band coincides with the end of hardening in the X-ray spectra. See Appendix for the discussion of significance of flares.

Fig.~\ref{opticalra} also shows the evolution of radio flux during outburst decay at different wavelengths from different observations. The green and red diamonds are from ATCA at 5.5 GHz and 9 GHz, respectively. We also showed the $H$ and $I$ band magnitudes to associate changes in radio to the changes in optical/NIR. Until MJD~55,320, the radio spectrum is optically thin. At MJD~55,311 and MJD~55,315, the radio core is detected with the VLBI, while the ATCA radio spectrum was still optically thin. This may indicate that there is contamination in the radio data, and the ATCA flux includes emission from a compact jet and some other interaction. The detection of the core corresponds to a little flare that can be seen in the $I$ band. Around MJD~55,335 the ATCA spectrum gets harder, and this is where the flux in $I$, $H$ band, and X-rays increase.

\begin {figure}[h]
\centering
\includegraphics[scale=0.8]{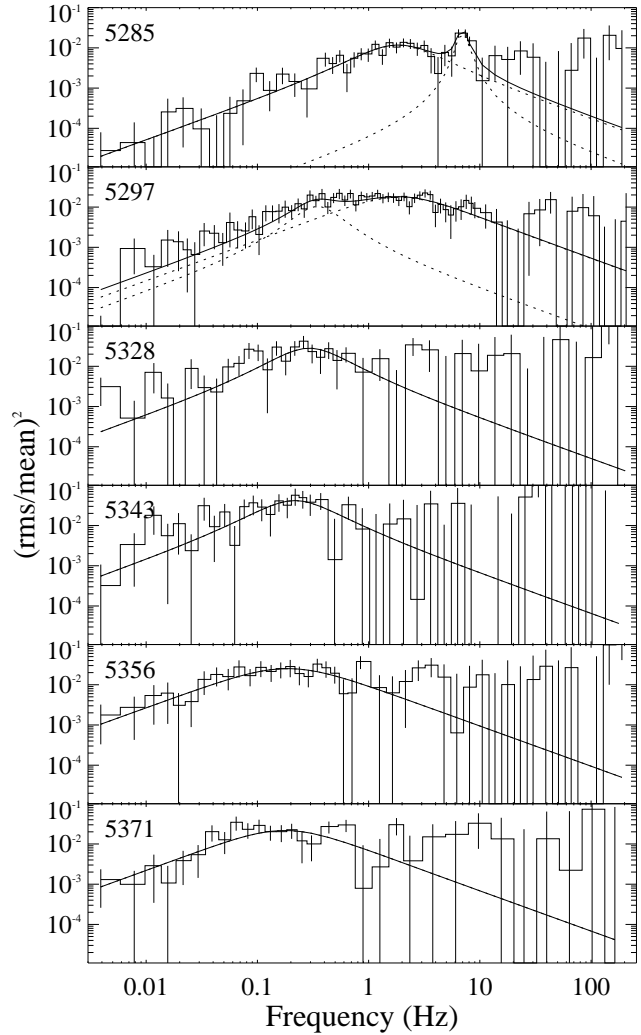}
\caption{Several selected PSDs from the intermediate state to low flux hard state. The vertical axes are frequency$\times$power and the dates (MJD-50,000) are indicated on the left of each panel. The peak of the Lorentzian gradually decreases.  The bottom three panels correspond to times before, during, and after the largest NIR flare, and there are no clear differences between these power spectra.} \label{timing1752}
\end{figure}

\begin{figure*}[t]
\epsscale{1.15}
\plottwo{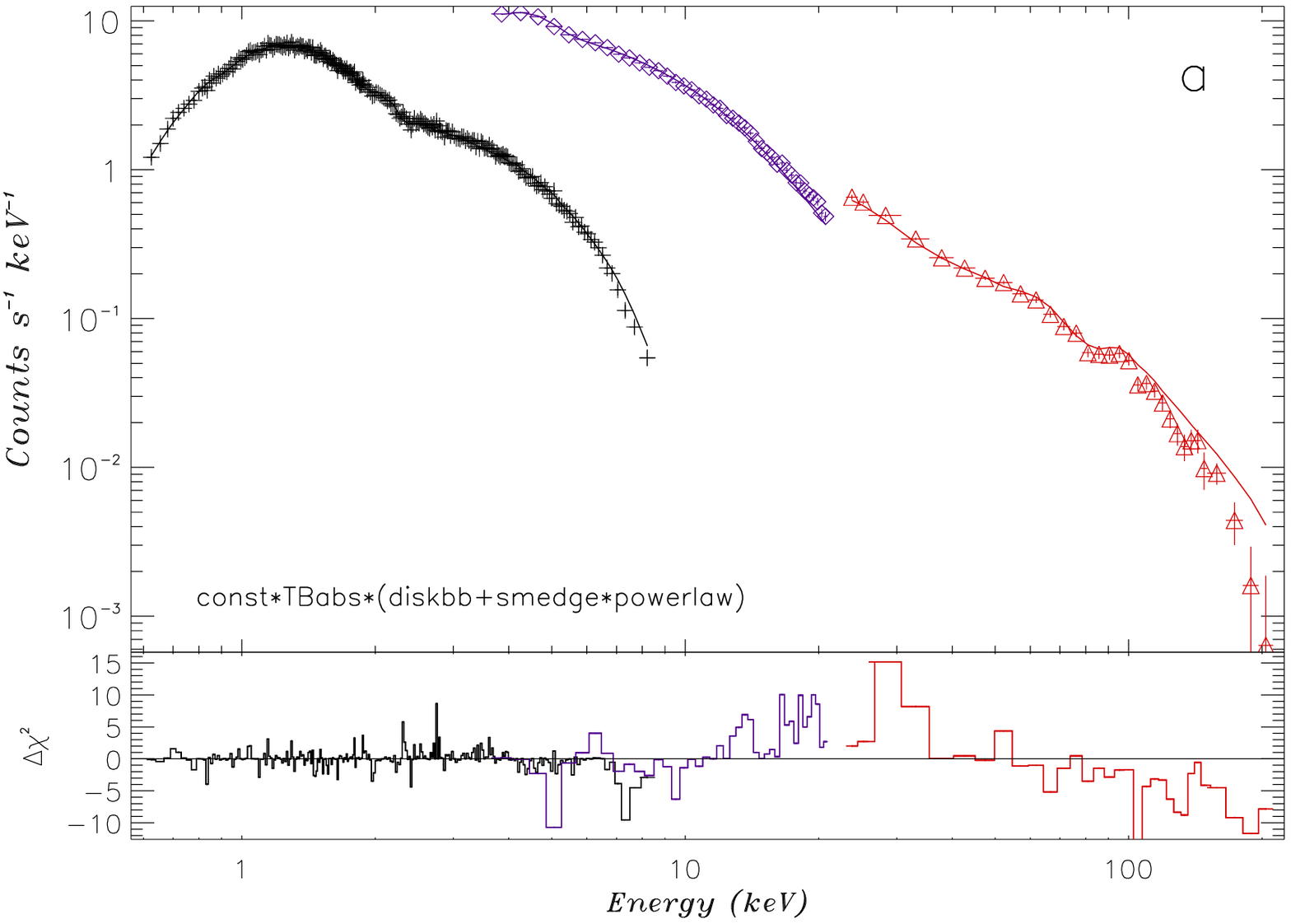}{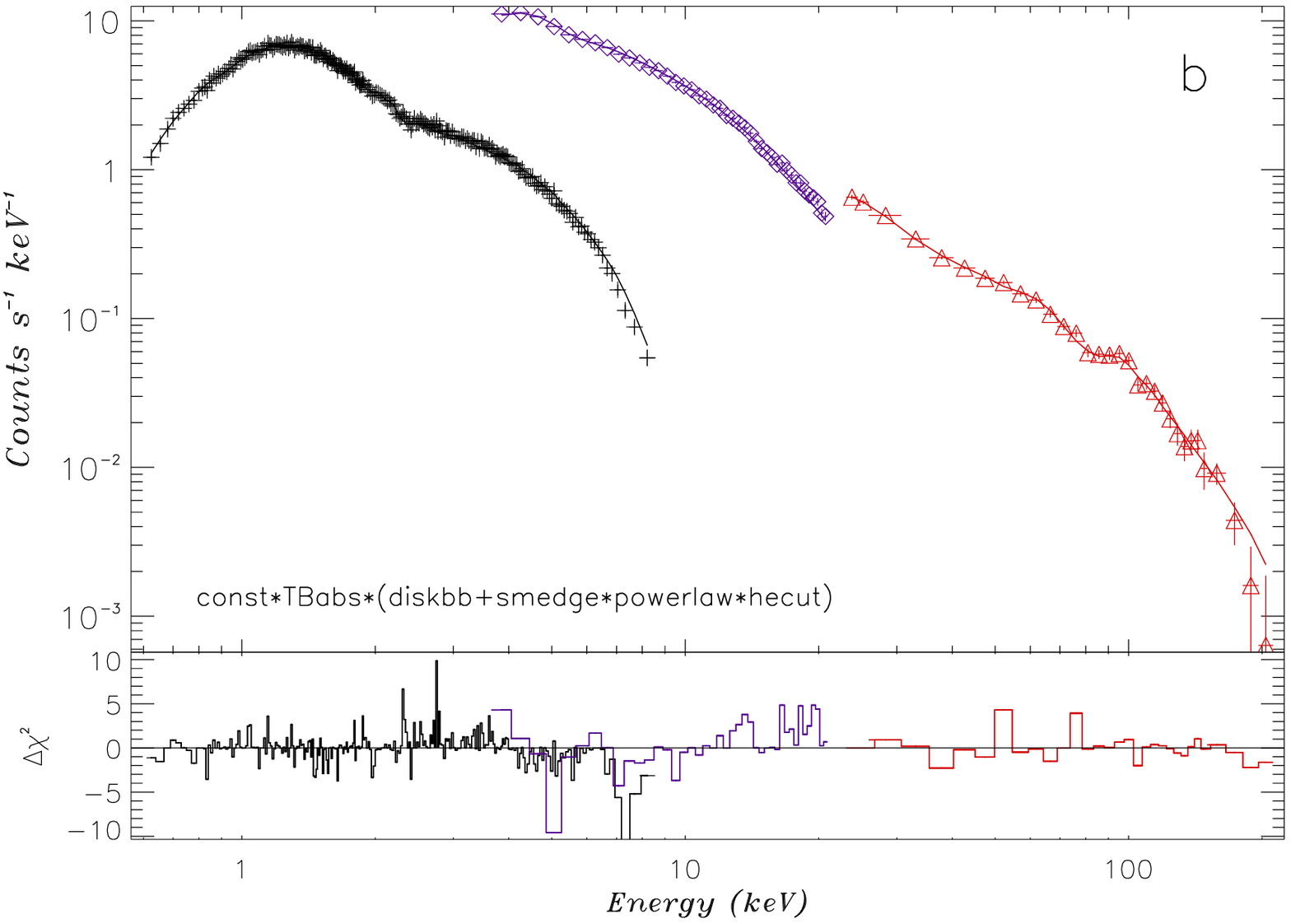}
\plottwo{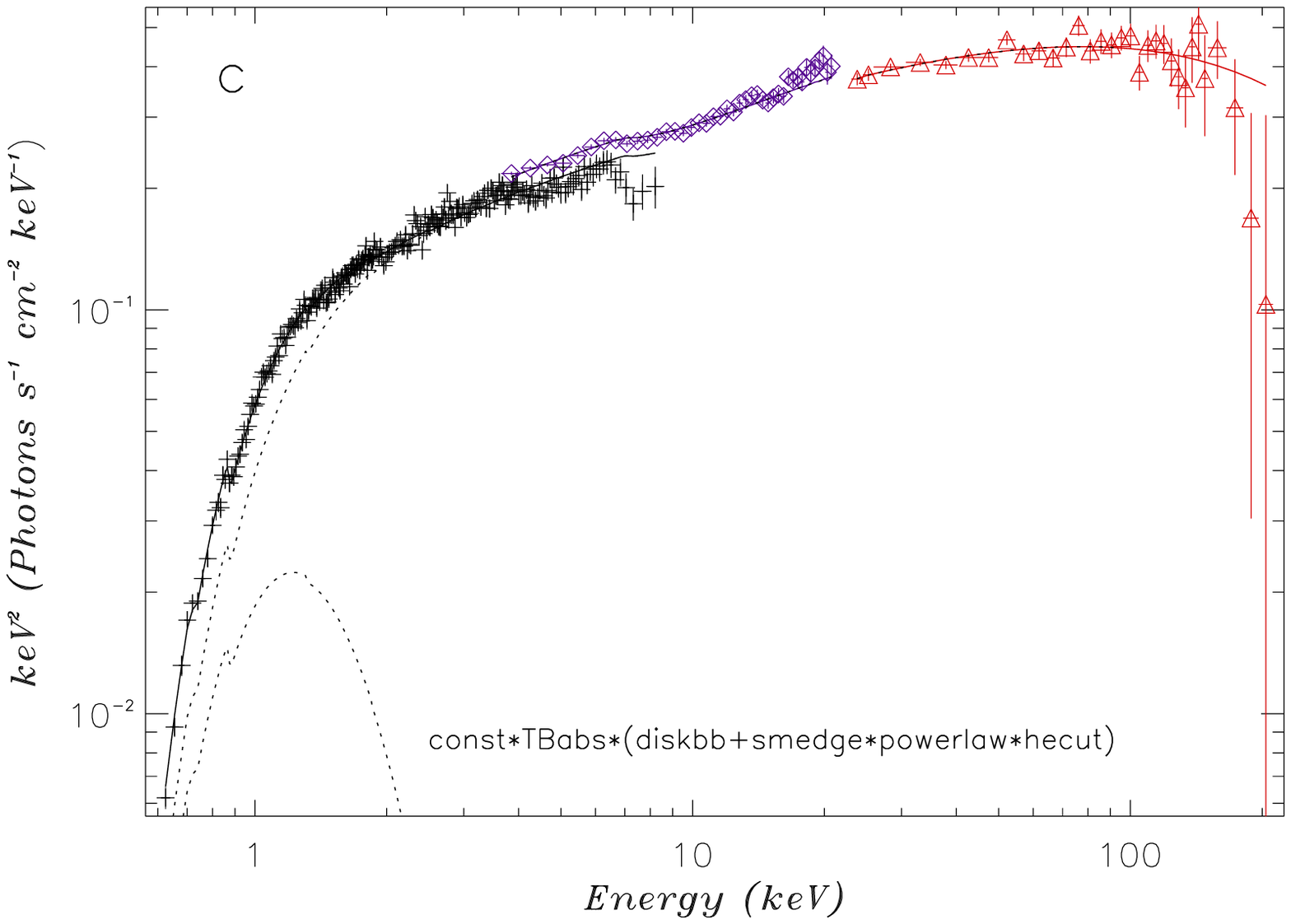}{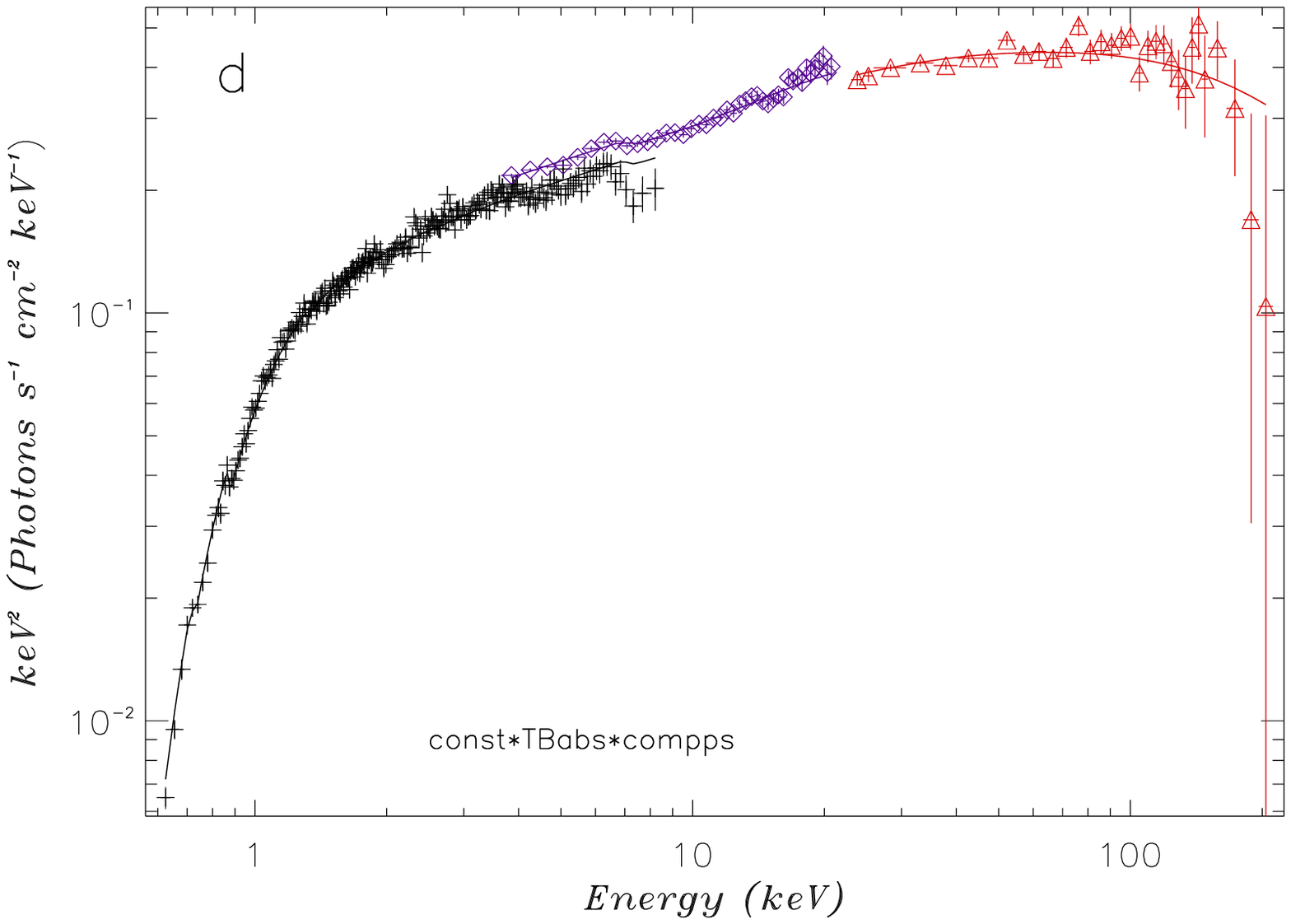}
\caption{Spectral fits to the \emph{Swift} XRT (shown with + sign, colored black in the online version), \emph{RXTE} PCA (shown with diamonds, colored purple in the online version) and \emph{INTEGRAL} ISGRI data (shown with triangles, colored red in the online version). a) Count spectrum and fit with a power-law and disk blackbody. b) Count spectrum and fit with a power-law and a high energy cut-off and disk blackbody. c) Unfolded spectrum with high energy cut-off. d) Unfolded spectrum with Comptonization model compps. A colored version of this figure is available in the online version.
} \label{combfit}
\end{figure*}

\begin{deluxetable*}{cccccccccc}

\tablecaption{The best fit parameters for simultaneous XRT, PCA, and ISGRI spectra}
\tablehead{
\colhead{Models} & \colhead{$N_{H}$}& \colhead{$kT_{bb}$, $T_{in}$\tablenotemark{a}} & \colhead{\emph{$\tau$}\tablenotemark{b}} & \colhead{$\nicefrac{\Omega}{2\pi}$\tablenotemark{c}} & \colhead{$kT_{e} (keV)$\tablenotemark{d}} & \colhead{$\Gamma$} & \colhead{$E_{cut} (keV)$\tablenotemark{e}} & \colhead{$E_{fold} (keV)$\tablenotemark{f}} & \colhead{$\chi^{2}/(dof)$} 
}
\startdata
CompPS+diskline & 0.60$_{-0.02}^{+0.02}$ & 0.28$_{-0.01}^{+0.01}$ & 1.24$_{-0.19}^{+0.24}$ & 0.53$_{-0.09}^{+0.11}$ & 105.2$_{-14.8}^{+14.8}$ &...&...&...& 312(323) \tabularnewline
power*highecut & 0.67$_{-0.04}^{+0.04}$ & 0.27$_{-0.02}^{+0.03}$&...&...&...& 1.67$_{-0.02}^{+0.02}$ & 25.4$_{-7.6}^{+7.3}$ & 236.5$_{-32.2}^{+41.8}$ & 342(323)\tabularnewline
\enddata
\tablenotetext{a}{The inner radius temperature of soft photons in multicolor disk.}\tablenotetext{b}{The vertical optical depth of the corona.}\tablenotetext{c}{The reflection factor.}\tablenotetext{d}{The electron temperature in the corona.}\tablenotetext{e}{The cut off energy.}\tablenotetext{f}{The folding energy.}
 \label{parameter}
\end{deluxetable*}

\subsection{Timing evolution}

As distinct multiwavelength changes occur during the brightening after MJD~55,340, we decided to check whether there are any changes in the PSD at the same time. The evolution of PSDs is shown in Fig.~\ref{timing1752}. We started from day MJD~55,285, during the first X-ray peak, and went down in flux to compare observations before and during the peak with similar fluxes. At MJD~55,285 two Lorentzians are required to fit the data, and the first Lorentzian peaks at $\sim$1.83 Hz. The rms amplitude of variability is $\sim$18.72\%. At MJD~55,297 the spectrum is now hard ($\Gamma$$\sim$1.7), and the rms amplitude is higher ($\sim$25.46\%). Still, two Lorentzians are required to fit the data, but the peak frequency of the first Lorentzian is shifted down to $\sim$0.35 Hz. For the rest shown in Fig.~\ref{timing1752}, a single Lorentzian is enough to fit the data. We note that the ridge correction significantly affects the rms amplitudes for flux levels less than $3 \times\ 10^{-10}$ ergs cm$^{-2}$ s$^{-1}$, 
which is reflected as large errors. At MJD~55,328 and MJD~55,356, the X-ray fluxes are almost the same with similar spectral parameters. The PSDs are also similar, and the rms and peak frequencies are consistent within errors. Likewise, the PSDs at MJD~55,343 and MJD~55,371 are also at similar flux levels, and show no significant change in spectral and timing properties (see Table~\ref{journal} for parameters). Even though the statistics are not good enough to demonstrate a significant difference (or lack thereof), the timing properties appear not to change before and during the flare, and similar X-ray spectral properties provide similar PSDs.

\begin{deluxetable*}{cccccc}[b]
  \tablecolumns{6} \tablewidth{360pt} \tablecaption{Resulting Hydrogen
    column density values obtained  from fits to individual absorption
    edges in the RGS data.  Values are in units of 10$^{22}$~cm$^{-2}$}
  \tablehead{  &  O  &  Fe  &  Ne  &  Mg  &  Average\tablenotemark{a}}
  \startdata
  ISM\tablenotemark{b}  & 1.04$\pm$0.05 & 0.88$\pm$0.04 & 1.17$\pm$0.12 & 1.08$\pm$0.18 & 0.96$\pm$0.03 \\
  Solar\tablenotemark{c}&    1.04$\pm$0.05     &    0.73$\pm$0.04    &
  1.17$\pm$0.12 & 0.99$\pm$0.14 & 0.88$\pm$0.03
\enddata
\tablenotetext{a}{Error weighted averages of the values found from individual edges.}
\tablenotetext{b}{As given by Wilms et al. (2000).}
\tablenotetext{c}{As presented by Asplund et al. (2009).}
\label{edges}
\end{deluxetable*}

\subsection{Broadband X-ray spectrum}\label{subsec:broadX}
 
To test whether spectral breaks disappear with the formation of jets, we conducted observations with the \emph{INTEGRAL} observatory. The ISGRI spectrum is combined with contemporaneous \emph{RXTE} PCA data. We started with our standard model, consisting of absorption, diskbb, smedge and power-law. We saw strong residuals indicating a high energy cut-off component in the fit (see Fig.~\ref{combfit}a). Adding a high energy cut-off (\emph{highecut}) significantly improved the fit (F-test results in a chance probability of 10$^{-20}$, see Fig.~\ref{combfit}b and c). The fit parameters can be found in Table~\ref{parameter}. 

Moreover, we also tried physical models and tried to establish whether an iron line is present in the spectrum. We fit the combined ISGRI, PCA, XRT spectrum first with the thermal Comptonization code, \emph{CompPS} (\citealt{Poutanen96}, see Fig.~\ref{combfit}d) and obtained a good fit with a reduced $\chi^{2}$ of 0.97. The statistical quality of the XRT spectrum was not good at the iron line region. We added \emph{diskline} \citep{Fabian89} to model the iron line emission at $\sim$6.4 keV. The improvement in the fit was not significant. Our dataset does not allow us to constrain the iron line parameters of this source at the time of the \emph{INTEGRAL} observation.

\begin{figure*}
\centering
\subfigure{\includegraphics[scale=0.4,angle=0]{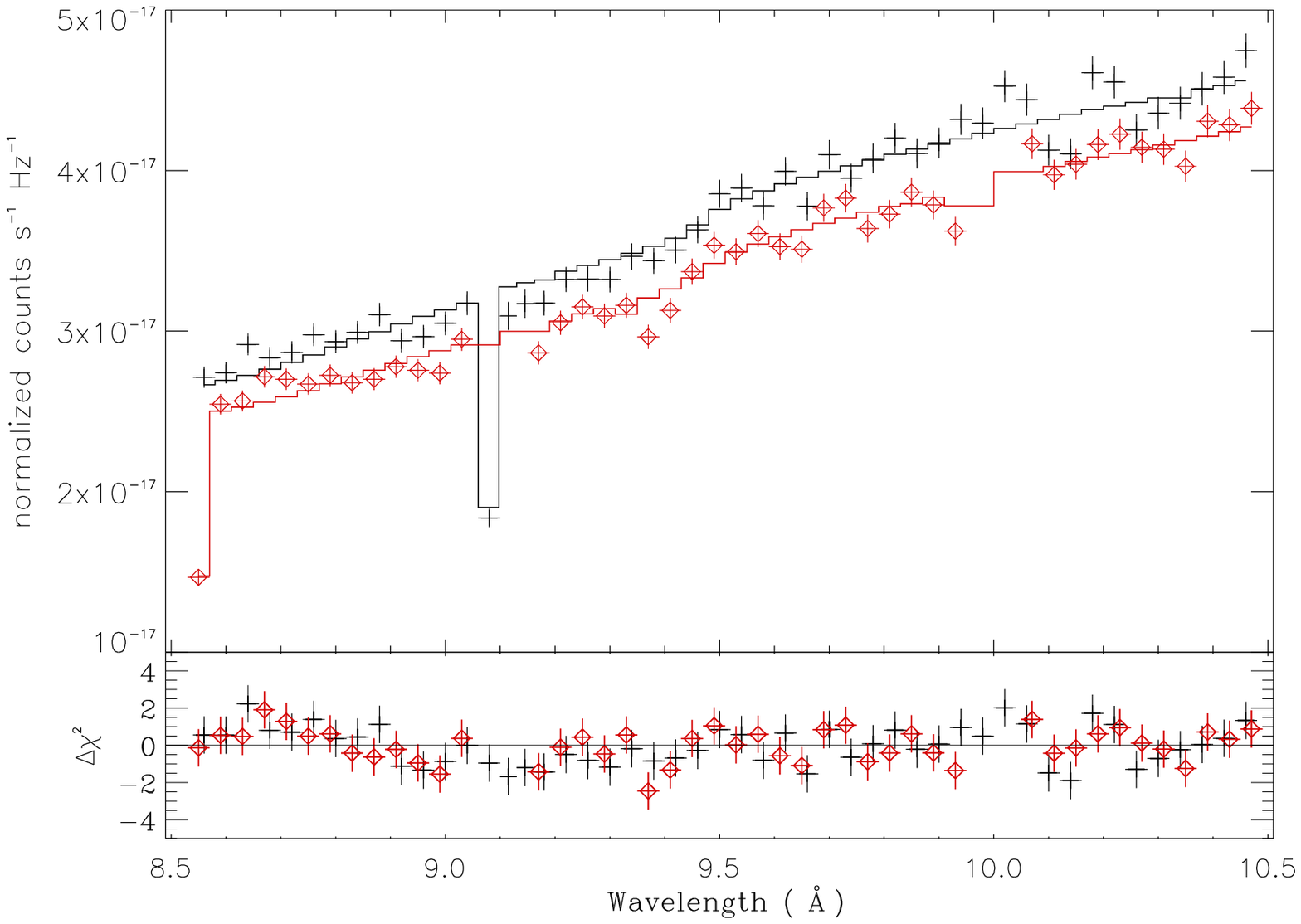}}
\subfigure{\includegraphics[scale=0.4,angle=0]{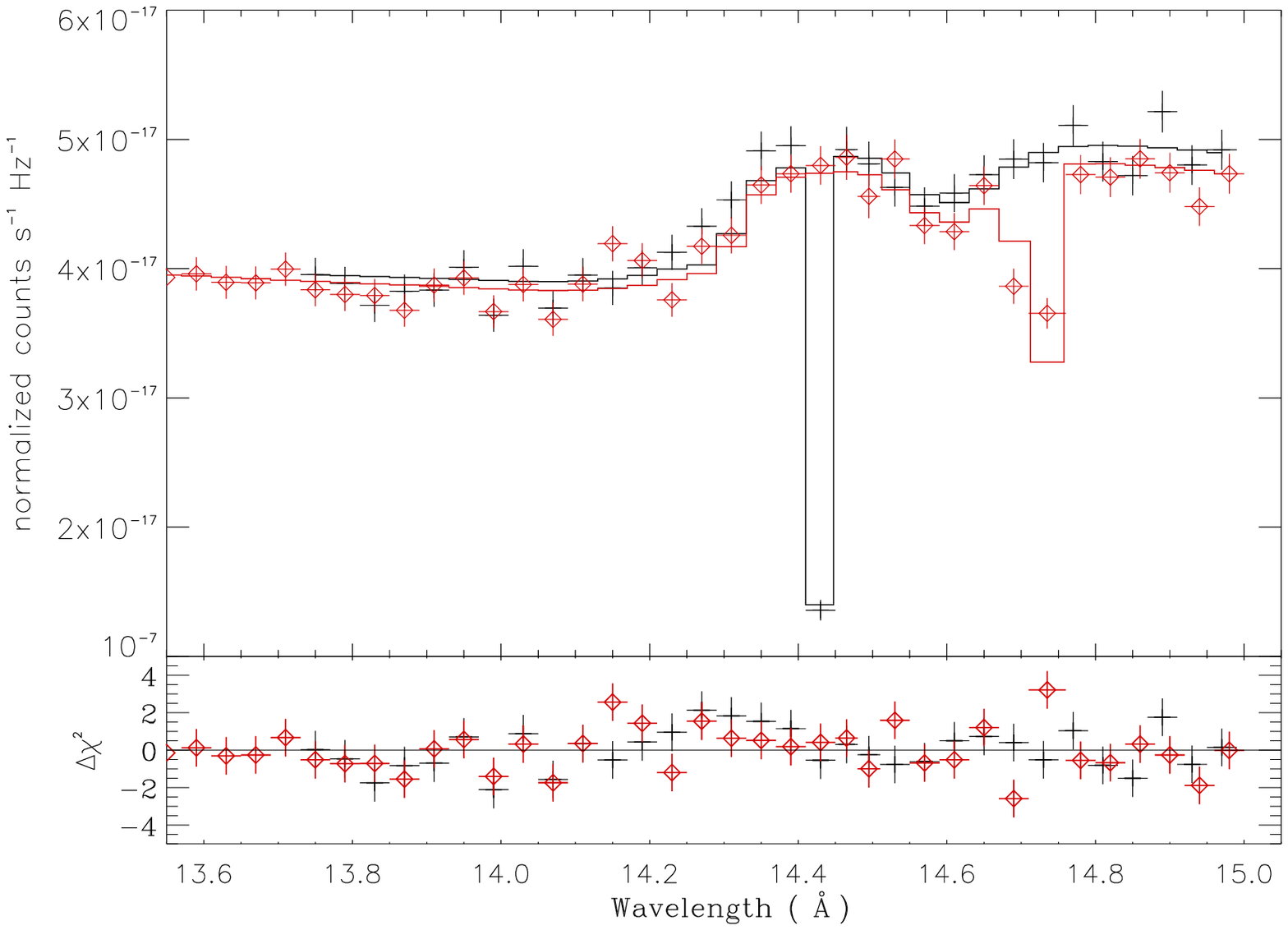}}
\centering
\subfigure{\includegraphics[scale=0.4,angle=0]{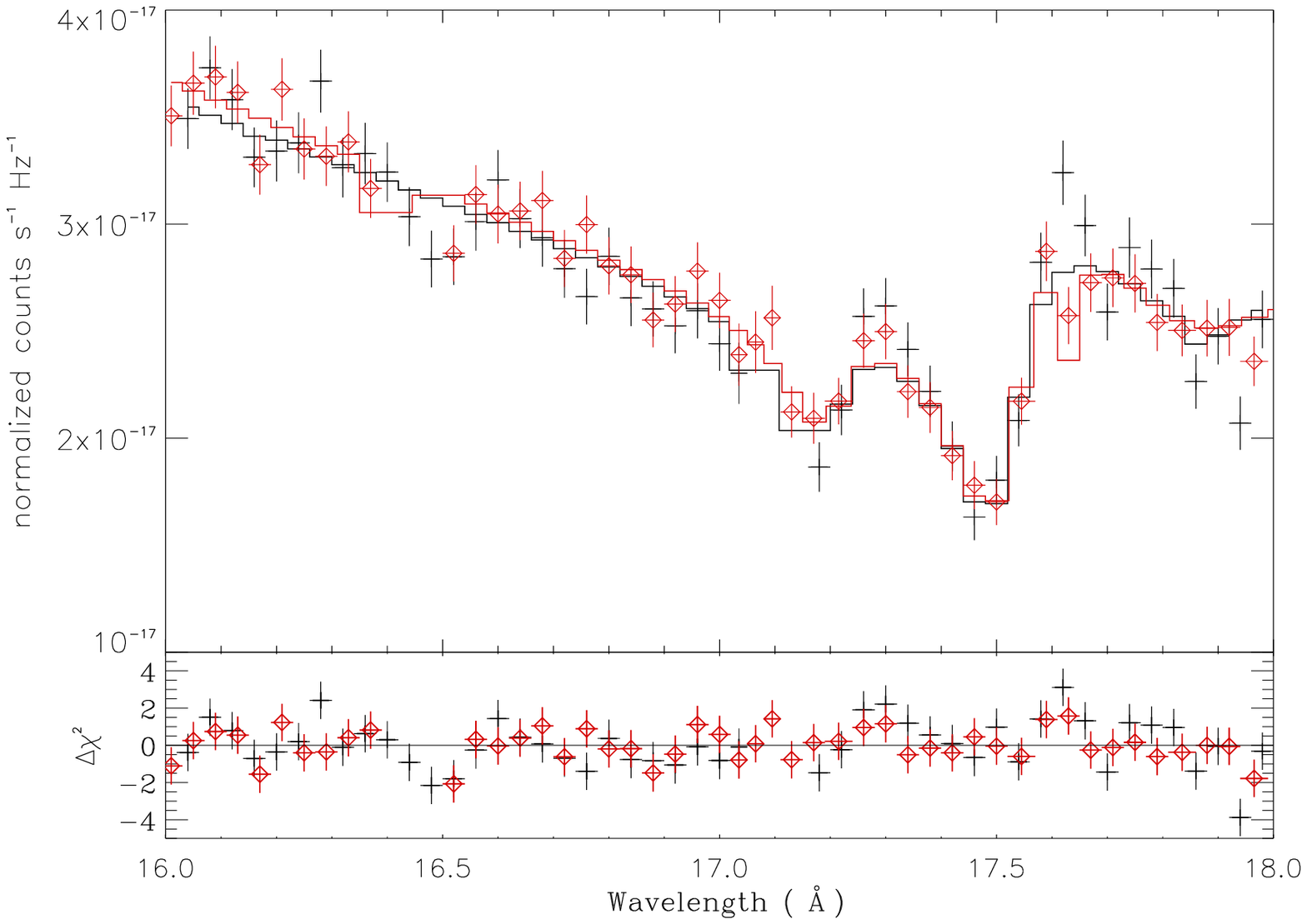}}
\subfigure{\includegraphics[scale=0.4,angle=0]{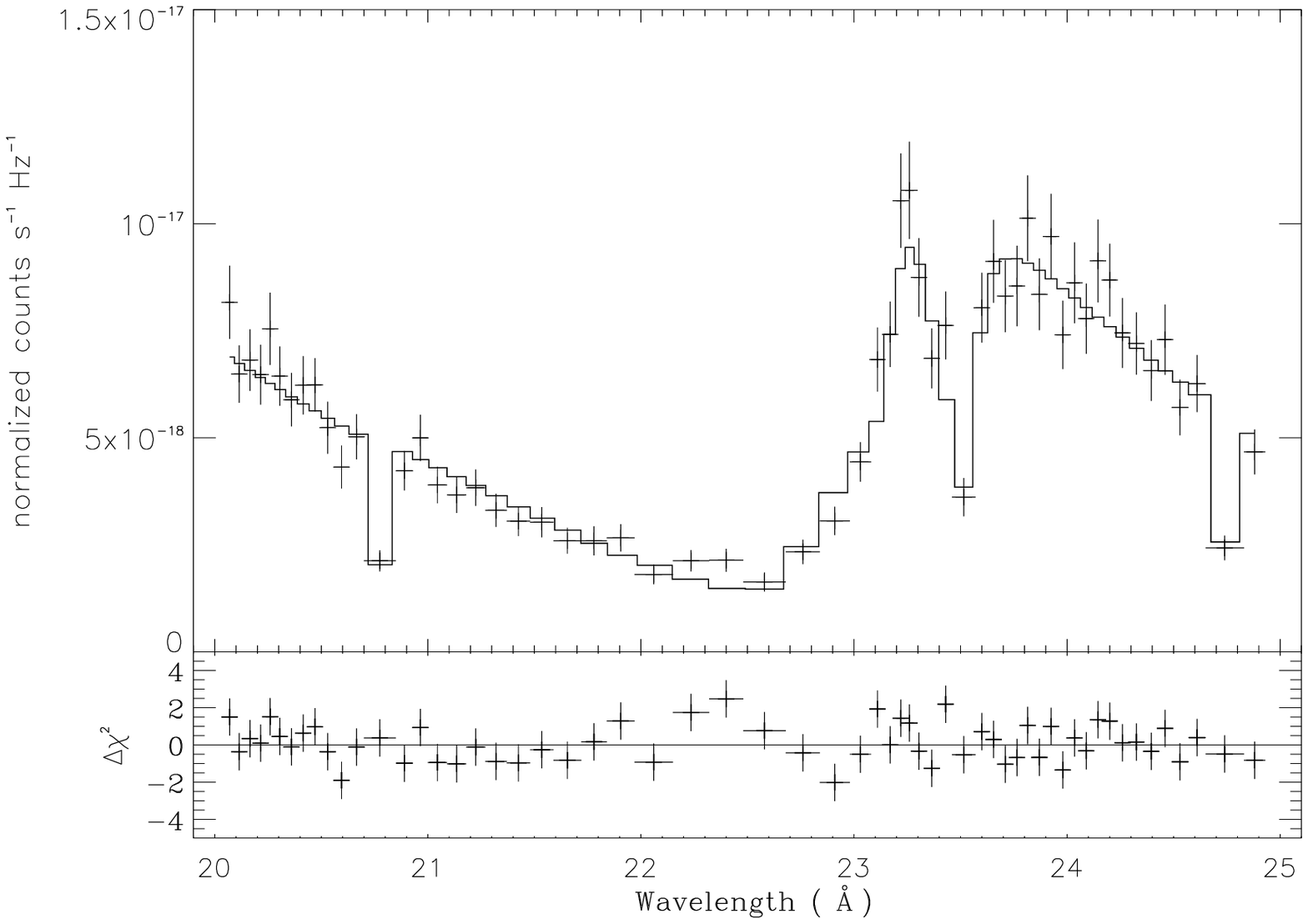}}
\caption{RGS1 (black) and  RGS2 (red) data at the  Mg (upper left), Ne (upper right), Fe (lower left), and O (lower right panel) absorption edge  regions are  shown in  the upper  panels of  each  plot.  Lower panels show the residuals from the best fit model. Sharp features are due to CCD gaps in the RGS detector. A colored version of this figure is available in the online version.}
\label{edge_figure}
\end{figure*}

\begin{figure*}
\centering
\subfigure{\includegraphics[angle=90,scale=0.35]{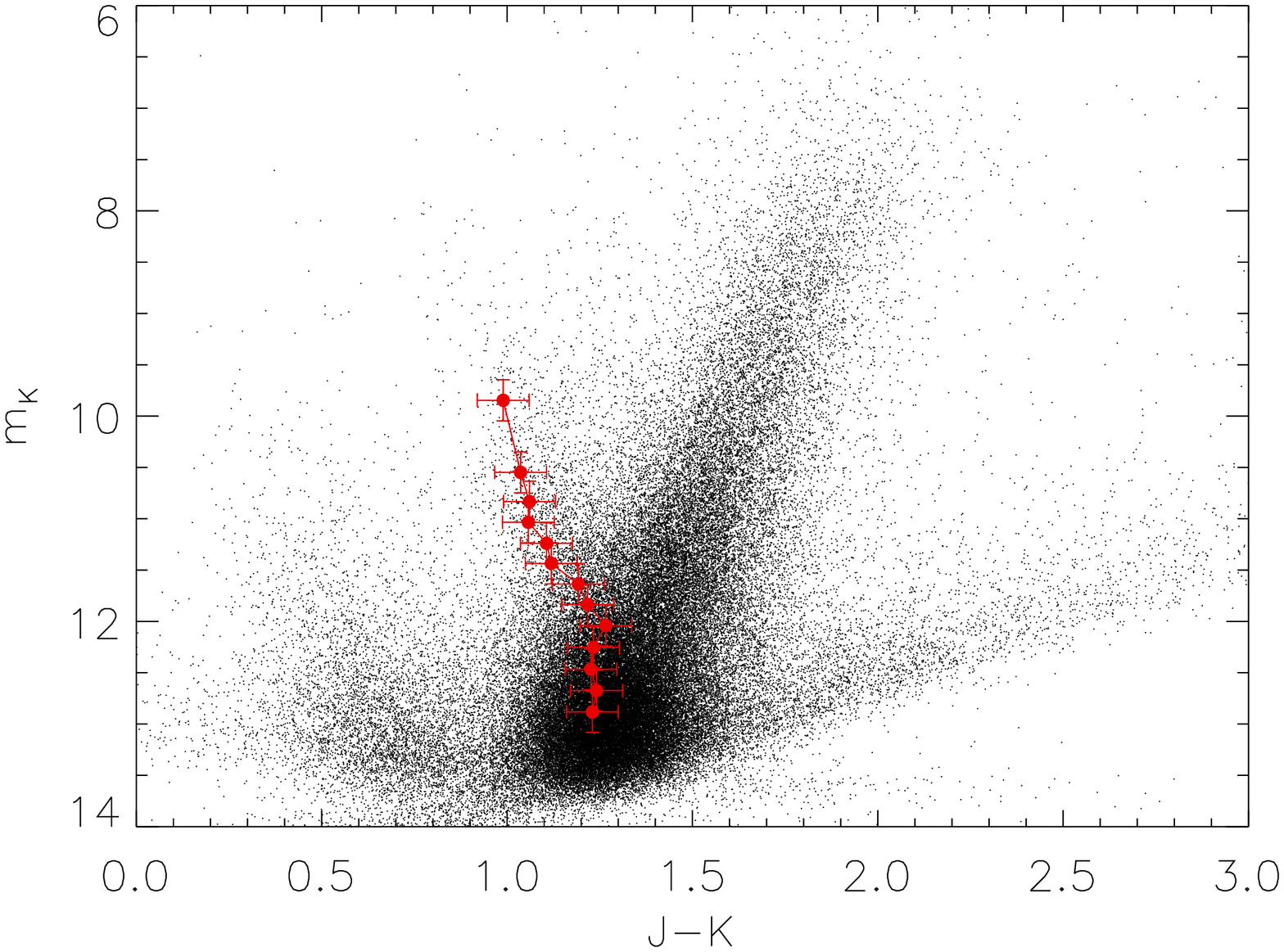}}
\subfigure{\includegraphics[scale=0.4]{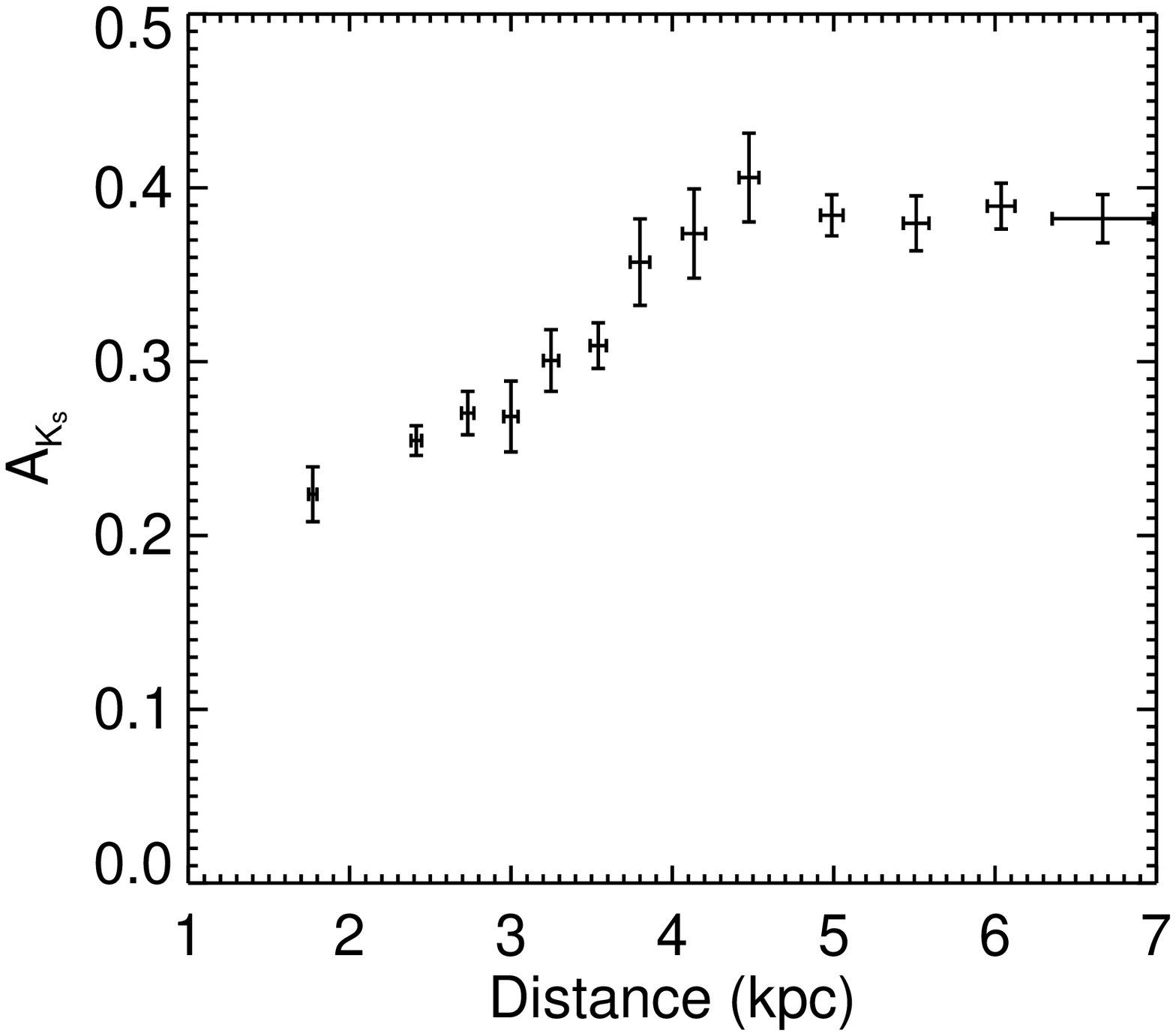}} 
\caption{Left: Near-IR color-magnitude diagram for the field centered around XTE J1752-223. Red crosses show points of maximum density of red clump stars for each individual magnitude bin and their uncertainties. Right: Evolution of near-IR extinction in the direction of XTE J1752-223 as derived from the 2MASS archive. A colored version of this figure is available in the online version.}
\centering
\label{dcm_xte}
\end{figure*}

\subsection{Distance to the source}\label{sec:distance}

Near-IR observations of red clump stars have been used to map the extinction as a function of distance along a given line of sight in the Galaxy, as reported in a number of studies by \cite{Paczynski98a,Lopez02,Cabrera05,Nishiyama06}. As they have a very narrow luminosity function (especially in the near-infrared), and they can be easily isolated from a color magnitude diagram, they have been largely used as standard candles in describing the geometry in the inner Galaxy \citep{Cabrera-Lavers07,Cabrera-Lavers08}.  For all the above, this population becomes a very powerful tool both for tracing and modeling the interstellar extinction in the more obscured areas of the Milky Way (see, e.g., \cite{Drimmel03} or \cite{Marshall06}).

These stars have also been used to estimate the distances of X-ray sources by comparing the estimates for the increase in extinction with the distance derived from the NIR data with the intrinsic extinction of the source derived from their X-ray data, with very successful results (see, e.g., \cite{Durant06,Castro08,Guver10}). Here, we follow a similar method to map extinction vs. distance in the direction of XTE J1752-223 by using photometric data from the 2MASS survey \citep{Skrutskie06}.

To find the intrinsic extinction in NIR, we first measured the interstellar X-ray absorption of the source in a way that is independent of assumptions about the intrinsic X-ray spectral properties by modeling individual absorption edges of the elements O, Ne, Fe, and Mg. For this purpose we only use the X-ray grating observation of the source obtained with the RGS on board \xmm.

To measure the Hydrogen column density towards \ST, we followed a method similar to \cite{Durant06} and \cite{Guver10}. We selected only a small wavelength range for each edge; 8.5$-$10.5~\AA~region for Mg, 13.5$-$15.0~\AA~for Ne, 16.0$-$18.0~\AA~for Fe, and 20.0-25.0~\AA~for the O edge (note that in this wavelength range, RGS2 has no sensitivity because of a CCD failure early in the mission.  Hence, we did not include data from RGS2 for the O edge). We assumed that the continuum in these small intervals can be modeled with a power-law function and modeled each edge using the {\it tbnew}\footnote{http://pulsar.sternwarte.uni-erlangen.de/wilms/research/tba-\\bs/} model \citep{Wilms12}. For spectral modeling we used XSPEC 12.7.0 \citep{Arnaud96}. We used the elemental photoelectric absorption cross-sections as given by \cite{Verner96}. Finally, we performed the fits both assuming ISM and solar abundances as given by \cite{Wilms00} and \cite{Asplund09}, respectively. We present the resulting Hydrogen column 
density values in Table~\ref{edges} for each element and show our fits to the data in Fig~\ref{edge_figure}.

In the O edge region (at 23.05 \AA) we noticed an absorption line like feature in the RGS1 spectrum that is not modeled in the {\it tbnew} model. We modelled this feature with a Gaussian to obtain a better fit for the absorption edge. We note that adding this component does not change the measured O column density although adding it improves our statistics. The position is consistent with S XIV resonance line. Similarly, we noticed a statistically significant absorption line like feature in the Ne edge region both in the RGS1 and RGS2 data at 14.6 \AA as well. Again we modelled this feature with a Gaussian to obtain a better fit for the continuum. This position is consistent with Fe XVIII resonance line and the line can be seen in Fig.~\ref{edge_figure}.

Following the method outlined in \cite{Guver10} and the relations presented by \cite{Guver09} and \cite{Cardelli89}, the Hydrogen column density we measure assuming an ISM abundance results in an optical extinction of A$_{\rm V}$=4.34$\pm$0.22 mag or near-infrared extinction of A$_{\rm K_{s}}$= 0.49$\pm$0.09 mag. Similarly assuming solar abundances, we get an optical extinction of A$_{\rm  V}$=3.98$\pm$0.21 or near-infrared extinction of A$_{\rm   K_{s}}$=0.46$\pm$0.08 mag. 

In Fig. \ref{dcm_xte} (right), we show the derived evolution of the extinction as we move toward the inner Galaxy in the direction of XTE J1752-223 (${\it l}$=6$^\circ$.42 , ${\it b}$=+2$^\circ$.11), while the region formed by the red clump stars can be seen in the color-magnitude diagram (see Fig. \ref{dcm_xte}, left). It can be seen that the extinction increases up to 4.5--5 kpc due to the interstellar dust in the Galactic plane in this direction, and remains nearly constant for distances larger than these. We observe this flattening since we have reached the bulge component at this distance along the line of sight, making it harder to accurately trace the extinction for larger distances. The red clump sources for the Bulge at a Galactic latitude of 2$^\circ$ above the plane completely dominates the star counts at this distance, and the (few) red clump sources for the Galactic disc are completely diluted. Hence,  extinction at greater distances is uncertain. Therefore, this method of using red clump stars, 
unfortunately, has limited capability in terms of measuring the distance to XTE J1752-223.  However, based on the extinction values we obtained for the source from X-ray spectroscopy and assuming that at least a significant part of the X-ray absorption is not due to intrinsic absorption, XTE J1752-223 cannot be closer than 5 kpc.

\section{Discussion}\label{sec:discussion}

\subsection{State transitions and jets}

The results of X-ray monitoring and the O/NIR light curves are shown in Fig.~\ref{evolution}, and the radio evolution is included in Fig.~\ref{opticalra} \citep[see also][]{Brocksopp13}.  Flare 1 in the $H$ band around MJD~55,280 is not significant, yet, it corresponds to a change in X-ray timing properties \citep{Buxton10atel}, indicating a transition to an intermediate state, as well as an increase in the $ATCA$ radio flux between MJD~55,275 and MJD~55,290 denoted as ``Peak 7'' in \cite{Brocksopp13}.  Close to the beginning of this flare, on MJD 55,277, the core is not detected with the $VLBI$ with an upper limit of 0.11 mJy. According to \cite{Brocksopp13}, the increase in the radio flux is probably from the interactions of jet components rather than compact jet turning on.

On the other hand, flare 2, between MJD~55,310 and MJD~55,330, is coincident with both the VLBI detection of the core on MJD~55,312 and a small increase of the ATCA flux. Since flare 2 also coincides with the end of hardening of X-ray spectra, it is more likely to be a result of compact jet forming \citep{Dincer12, Kalemci12}, supporting the conclusion of \cite{Brocksopp13}. Unfortunately, the lack of $VLBI$ coverage between MJD~55,277 and MJD~55,312 prevents us from coming to strong conclusions on when exactly the compact core turned on.

The $ATCA$ spectral index increases towards MJD~55,350, and the spectrum becomes consistent with a flat spectrum on MJD~55,350 {\bf{(hardness/intensity at the time are shown in Fig.~\ref{fig:hid}).}} This spectral evolution coincides with the flare 3 seen in both the $I$ and the $H$ band starting at MJD~55,335. A compact jet may be dominating the O/NIR and radio emission after this date \citep[also see][]{Russell12}. An increase in O/NIR flux along with a radio spectral evolution from an optically thin to optically thick emission is also seen for GX 339-4 \citep{Corbel12}. 
	
If the emission from a compact jet dominates the X-ray emission during the second long flare between MJD~55,340 and 55,380 \citep{Russell12}, one could expect some change in the timing properties before and after the peak.  We did not observe a significant change in the PSDs, their rms amplitudes, and the peak frequencies when the compact jet is present. However, because the errors of the peak frequencies in these datasets are not negligible, $\sim$30\%, timing studies for brighter sources might be more useful to discuss the effects of jets on timing.

\subsection{Spectral breaks in hard X-rays}

The presence of spectral breaks is well established during the transition to (or from) the hard state \citep{Kalemci05, Motta09}. There is some indication that the breaks disappear after the jet turns on  based on spectral fitting of data from HEXTE \citep{Kalemci06_mqw}. However, there is no conclusive evidence that relates the jet formation to the evolution of spectral breaks as the short monitoring observations with \rxte\ often lack the statistics required to characterize this evolution. To obtain better statistics in hard X-rays, we triggered our \emph{INTEGRAL} observation around the time that the spectral index is hardest. The observation took place just before the core is detected wıth the $VLBI$. The broadband combined spectrum clearly shows a break, with a folding energy at $\sim$237 keV. This result is consistent with earlier work showing that a cut-off is present before a compact jet dominates the radio and O/NIR emission.  Unfortunately, the timing of the observation did not allow us to test the 
hypothesis that the cut-offs disappear after the strong compact jets dominates the multiwavelength emission. The hard X-ray spectrum of the source is consistent with thermal Comptonization.  

\subsection{Distance}

The method using red clump stars provides a lower limit to the distance to the source of $>$ 5 kpc. \cite{Shaposhnikov10} estimate the distance of the source as about 3.5 kpc. At this distance, the Galactic extinction is about 60\% of what we infer from the absorption edges observed with \xmm\ RGS. Assuming that the intrinsic absorption in this source is not as high as 2 magnitudes in the optical, this distance estimate is incompatible with our results. Such a high intrinsic absorption seems hard to understand given that no dips in the X-rays have been observed, which could have been interpreted as neutral matter absorbing X-rays emitted from the boundary layer. It has been found by \cite{Miller09} that the X-ray absorption as derived from photoelectric absorption edges remains constant as the luminosity and spectral states of X-ray binaries, including black hole systems, vary. This finding suggests that the ISM strongly dominates the measured neutral Hydrogen column density in the spectra of X-ray binaries.

Also, possible uncertainties in the extinction law used for deriving the interstellar extinction (see, e.g., \citealt{Nishiyama09, Gonzalez12}) cannot support such high differences from the estimate by \cite{Shaposhnikov10}, as the affects of such differences are within the range covered by the errors shown in  Fig.~\ref{dcm_xte}. Hence, the assumption of XTE J1752-223 being no closer than 5 kpc is not produced by uncertainties in the NIR extinction measurements.

We note that the distance of 3.5 kpc is also not compatible with the overall behavior of GBHTs in terms of their transition luminosities \citep{Maccarone03_b,Ratti12}. As shown in \cite{Kalemci12}, the overall luminosity evolution of this source becomes compatible with other black hole transients if we assume a distance around 8 kpc, and a distance of 3.5 kpc would make the behavior of this source quite different from the other black hole binary systems. Future observations of this source during a new outburst can test our measurements and provide clues about the any variabilities in the amount of X-ray absorption, which would be a direct evidence for intrinsic absorption.

Apart from the evidence from the transition luminosities, the radio - X-ray correlation is also consistent with the behavior of other radio-quiet sources if the distance is 8 kpc \citep{Brocksopp13}. The larger distance not only boosts the minimum power in the jet, but also increases the speed of the jet. As indicated by \cite{Yang11}, a larger distance means that the XTE~J1752$-$223 may be a superluminal source.

\section{Summary and Conclusions}

In this work, we analyzed data from the Galactic black hole binary XTE~J1752$-$223 from the radio band all the way to hard X-rays during its outburst decay in 2010. We investigated the evolution of X-ray spectral and temporal properties of the source using \rxte\ and \emph{Swift} and compared the results to the evolution of fluxes in the $I$ and $H$ band \emph{SMARTS} data and also to the spectral evolution of the $ATCA$ radio data. We also studied the $XMM$-$Newton$ RGS data to find the extinction towards the source. The important results from this analysis are summarized below:

\begin{itemize}

\item We showed that the detection of the radio core with the $VLBI$ corresponds to a small flare in the $I$ band flux (flare 2 in Fig.\ref{opticalra}). This flare is followed by a larger flare in the $I$ and $H$ band flux  (flare 3 in Fig.\ref{opticalra}).
\item Flare 3 in the $I$ and the $H$ band coincides with the $ATCA$ radio data becoming optically thick, relating the compact jet to the O/NIR changes.
\item The short term X-ray timing properties do not show a significant change during flares in the O/NIR.
\item The broadband X-ray spectrum including $INTEGRAL$ ISGRI indicates a high energy cut-off, which is also seen in other sources before strong compact jets are observed. The spectrum is consistent with thermal Comptonization.
\item We showed that the distance to the source cannot be less than 5 kpc. Based on the analysis of transition luminosities and also the behavior of of radio - X-ray flux correlation, the distance is more likely to be close to 8 kpc.  

\end{itemize}

\acknowledgments 

YYC, EK, TD and SC acknowledge support from FP7 Initial Training Network Black Hole Universe, ITN 215212. EK and TD acknowledges T\"{U}B\.{I}TAK grant 111T222. JAT acknowledges partial support from NASA {\em Swift} Guest  Observer grant NNX10AK36G and also from the NASA Astrophysics  Data Analysis Program grant NNX11AF84G. EK thanks Sinan Alis for his help with IRAF analysis. The authors thank all scientists who contributed to the T\"{u}bingen Timing Tools and Craig Markwardt for MPFIT algorithms. This publication makes use of data products from the Two Micron All Sky Survey, which is a joint project of the University of Massachusetts and the Infrared Processing and Analysis Center/California Institute of Technology, funded by the National Aeronautics and Space Administration and the National Science Foundation. The Australia Telescope is funded by the Commonwealth of Australia for operation as a National Facility managed by CSIRO. 




\clearpage

\appendix

\section{Significance of flares}

\begin {figure}[t]
	\centering
	\includegraphics[scale=0.75]{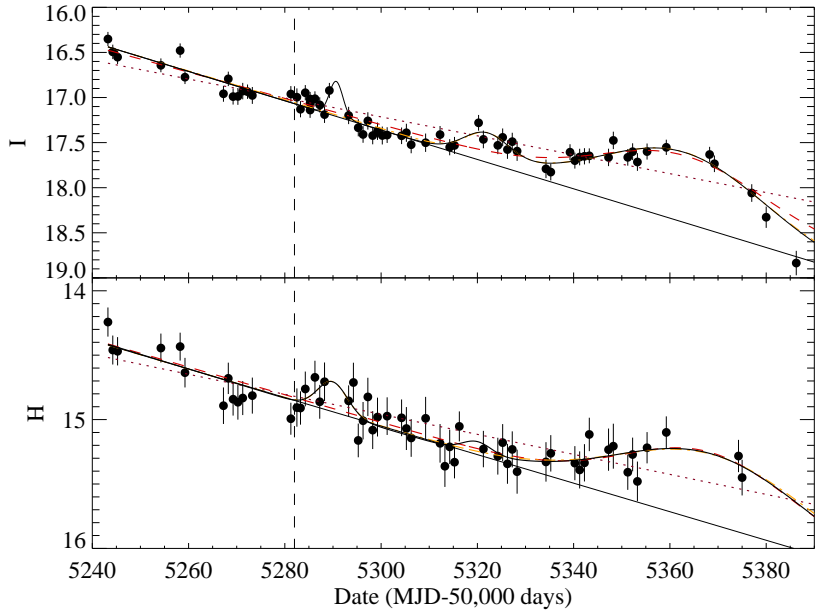}
	\caption{Straight line + Gaussian fits to the flares discussed in the article. Top panel shows the $I$ band light curve and the bottom panel shows the $H$ band light curve. The dotted lines (colored purple in the online version) are linear fits to the data.  The long dashed curves (colored red in the online version) are the fits with straight line+Gaussian for the flare 3. The dotted-dashed curves (colored orange in the online version) are fits with straight line + 2 Gaussians. The solid curves (colored black in the online version) are the overall fits with straight line + 3 Gaussians. The solid, straight line represents the linear fits to the continuum when all Gaussians are present. A colored version of this figure is available in the online version.} \label{fig:gauss}
\end{figure}

We have discussed three flares shown in Fig.~\ref{opticalra}. The first flare was the subject of \cite{Buxton10atel}, and the third flare was discussed in detail in \cite{Russell12}. Close to the dates of the detection of the radio core, we realized that there is some excess emission especially in the $I$ band. We then tested the significance of all flares in the $I$ and the $H$ band. To do that, we assumed that there is underlying emission that exponentially decays, which would correspond to a linear decrease in magnitudes \citep{Buxton04, Kalemci05, Russell10, Buxton12, Dincer12}. We first fitted the entire datasets
 with straight lines (dotted lines in Fig.~\ref{fig:gauss}). For both bands the fits were very poor (the $\chi^{2}$ values of all the fits are given in Table~\ref{table:sig}). Following \cite{Buxton04}, we then fitted the excesses over the continuum with Gaussians.

We started with the most obvious excess on \wsim MJD~55,370 (flare 3, long dashed lines in  Fig.~\ref{fig:gauss}). This flare is required in both bands. Some residuals were present in the $I$ band around MJD~55,320 and around MJD~55,290 in the $H$ band. Adding Gaussians in the fit to these flares (dotted-dashed lines in Fig.~\ref{fig:gauss}) we found that flare 2 is statistically significant in the $I$ band whereas flare 1 in the $H$ band is not statistically significant. Finally, for completeness, we added Gaussians around MJD~55,290 in the $I$ band and around MJD~55,320 in the $H$ band and refit the datasets (solid curves in  Fig.~\ref{fig:gauss}). These Gaussians are not formally required in the fit (chance probabilities obtained using the $F-test$ are given in Table~\ref{table:sig}).

\begin{deluxetable}{cccccc}[b]
\tablewidth{0pt}
\tablecaption{Fit results to the O/NIR light curves}
\tablehead{
\colhead{Flare \#}  & \colhead{$\chi^{2}_{1}$\tablenotemark{a}} & \colhead{$DOF_{1}$\tablenotemark{b}} &  \colhead{$\chi^{2}_{2}$\tablenotemark{c}} & \colhead{$DOF_{2}$\tablenotemark{d}}  & \colhead{$P$\tablenotemark{e}}
}
\startdata
\multicolumn{6}{c}{Fits to $I$-band light curve.} \tabularnewline 
3 & 163.40     &    62    &   80.72    &     59  &    4.14$10^{-9}$ \tabularnewline
2 &  80.72    &     59     &  58.77     &     56  &  4.56$10^{-4}$ \tabularnewline
1 &  58.77   &    56    &   50.44    &     53  &   0.043  \tabularnewline
\multicolumn{6}{c}{Fits to $H$-band light curve.} \tabularnewline 
3 &  63.72     &   55     & 40.41      &     52  &   2.62$10^{-5}$     \tabularnewline
1 & 40.41     &   52     & 35.91      &     49  &   0.12 \tabularnewline
2 &  35.91   &   49    &  35.50     &     46  &   0.91\tablenotemark{f} 
\enddata
\tablenotetext{a}{$\chi^{2}$ before adding the new component.}
\tablenotetext{b}{Degrees of freedom before adding the new component.}
\tablenotetext{c}{$\chi^{2}$ after adding the new component.}
\tablenotetext{d}{Degrees of freedom after adding the new component.}
\tablenotetext{e}{Chance probability from the $F$-distribution.}
\tablenotetext{f}{To get a fit, the width of the Gaussian was limited to 3-5 days.}
 \label{table:sig}
\end{deluxetable}

      



\end{document}